\documentclass{aastex63}

\received{September 2, 2020}
\revised{December 31, 2020, March 23, 2021, April 05, 2021}
\accepted{April 12, 2021}
\submitjournal{AJ}

\shorttitle{Aerosol Measurement using Wide-field Photometry}
\shortauthors{Ebr et al.}

\begin{document}

\title{A New Method for Aerosol Measurement using Wide-field Photometry}

\correspondingauthor{Jan Ebr}
\email{ebr@fzu.cz}

\author[0000-0001-8807-6162]{Jan Ebr}
\affiliation{FZU -- Institute of Physics of the Czech Academy of Sciences, Na Slovance 1999/2, Prague 182 21, Czech Republic}

\author[0000-0003-0035-651X]{Sergey Karpov}
\affiliation{FZU -- Institute of Physics of the Czech Academy of Sciences, Na Slovance 1999/2, Prague 182 21, Czech Republic}

\author[0000-0001-5341-6008]{Ji\v{r}\'{i} Eli\'{a}\v{s}ek}
\affiliation{FZU -- Institute of Physics of the Czech Academy of Sciences, Na Slovance 1999/2, Prague 182 21, Czech Republic}
\affiliation{Institute of Theoretical Physics, Faculty of Mathematics and Physics, Charles University,\\ V~Hole\v{s}ovi\v{c}k\'ach 2, 180\,00 Prague, Czech Republic}

\author[0000-0002-5870-8947]{Ji\v{r}\'{i} Bla\v{z}ek}
\affiliation{FZU -- Institute of Physics of the Czech Academy of Sciences, Na Slovance 1999/2, Prague 182 21, Czech Republic}

\author{Ronan Cunniffe}
\affiliation{FZU -- Institute of Physics of the Czech Academy of Sciences, Na Slovance 1999/2, Prague 182 21, Czech Republic}

\author[0000-0003-1858-5652]{Ivana Ebrov\'{a}}
\affiliation{Nicolaus Copernicus Astronomical Center, Polish Academy of Sciences, Bartycka 18, 00-716 Warsaw, Poland}

\author{Petr Jane\v{c}ek}
\affiliation{FZU -- Institute of Physics of the Czech Academy of Sciences, Na Slovance 1999/2, Prague 182 21, Czech Republic}

\author[0000-0003-3922-7416]{Martin Jel\'{i}nek}
\affiliation{Astronomical Institute of the Czech Academy of Sciences, Fri\v{c}ova 298, Ond\v{r}ejov 251 65, Czech Republic}

\author[0000-0002-3130-4168]{Jakub Jury\v{s}ek}
\affiliation{FZU -- Institute of Physics of the Czech Academy of Sciences, Na Slovance 1999/2, Prague 182 21, Czech Republic}

\author[0000-0001-7748-7468]{Du\v{s}an Mand\'{a}t}
\affiliation{FZU -- Institute of Physics of the Czech Academy of Sciences, Na Slovance 1999/2, Prague 182 21, Czech Republic}

\author[0000-0002-0967-0006]{Martin Ma\v{s}ek}
\affiliation{FZU -- Institute of Physics of the Czech Academy of Sciences, Na Slovance 1999/2, Prague 182 21, Czech Republic}

\author[0000-0002-8421-0456]{Miroslav Pech}
\affiliation{FZU -- Institute of Physics of the Czech Academy of Sciences, Na Slovance 1999/2, Prague 182 21, Czech Republic}

\author[0000-0002-3238-9597]{Michael Prouza}
\affiliation{FZU -- Institute of Physics of the Czech Academy of Sciences, Na Slovance 1999/2, Prague 182 21, Czech Republic}

\author[0000-0002-1655-9584]{Petr Tr\'{a}vn\'{i}\v{c}ek}
\affiliation{FZU -- Institute of Physics of the Czech Academy of Sciences, Na Slovance 1999/2, Prague 182 21, Czech Republic}



\begin{abstract}

We present a new method to measure the vertical aerosol optical depth (VAOD) during clear nights using a wide-field imager -- a CCD camera with a photographic lens on an equatorial mount. A series of 30-second exposures taken at different altitudes above the horizon can be used to measure the VAOD with a precision better than 0.008 optical depths within a few minutes. Such a measurement does not produce any light and is thus suitable for use at sites where other astronomical instruments are located. The precision of the VAOD measurement depends on laboratory calibration of spectral properties of the system and of the response of the camera electronics to varying illumination levels, as well as careful considerations of details of stellar photometry and modelling of the dependence of measured stellar fluxes star color and position within the field of view. The results obtained with robotic setups at the future sites of the Cherenkov Telescope Array show good internal consistency and agreement with the simultaneous measurements from a Sun/Moon Photometer located at the same site.

\end{abstract}



\section{Introduction} \label{sec:intro}

The observation of natural light sources outside of the atmosphere allow for the measurement of the integral optical depth in the direction of the source without production of any artificial light (as in the case of laser-based methods). Subtracting the contribution of molecular scattering and absorption, which can be derived from meteorological data, the aerosol optical depth can be obtained. The Sun, as a bright and stable source, has been commonly used for this purpose and photometers have become a standard tool in aerosol measurements during daytime \citep{sunphot}; more recently, the Moon has been used in the same way to extend the measurements into at least some nights, with some limitations on precision due to limited photometer sensitivity and difficulties in precisely describing the variations in lunar brightness \citep{moonphot}. Using stars as reference sources extends the measurement capabilities to nights when the Moon is not visible as well as to arbitrary positions on the sky, but brings its own set of challenges. The stars, being much fainter than the Sun or the Moon, require a more sensitive optical detector, which must be kept properly calibrated \citep{starphot1, starphot2}. Moreover, if the ability to measure the optical depth in an arbitrary direction is required, the problem of insufficient accuracy of available stellar photometric catalogues becomes apparent.

The need for an absolute calibrated instrument can be avoided using the well-known Langley approach \citep{langley} which relies on the known dependence of optical depth on the zenith angle for the case of horizontally uniform (stratified) atmosphere and allows the simultaneous determination of the calibration constant of the instrument and vertical optical depth (VOD) $\tau$ by observing a star at different zenith angles, that is, at different values of airmass $A$ (which is defined here as relative to the airmass in the vertical direction). The main limitation of this method when using a single celestial target, such as the Sun, Moon, single reference star or field of stars, is that VOD variations occur much more quickly (minutes) than the Earth's rotation causes the chosen reference source to traverse a large span in zenith angles (hours). Thus, we propose a method where we measure a large number of stars at a large span of airmasses in a short amount of time (a few minutes) using a series of images (which we call an ``altitude scan'') taken with a wide-field imager. The VOD, and, after the subtraction of the molecular components, the vertical aerosol optical depth (VAOD), is then extracted from a fit as a function of the airmass giving the difference between the measured and expected brightness of the stars.  The expected brightness is based on a homogeneous all-sky catalogue and a calibrated model of the optics and the detector (which can also be derived from the same data). Note that this value is sometimes referred to in literature simply as the aerosol optical depth (AOD) and the dependence on airmass is implied.

The method to determine the VAOD from the airmass-dependence of extinction is in principle straightforward as the latter is expressed by a known analytical formula, but a significant instrument calibration effort is necessary to reach a precision better than 0.01 optical depths (hereafter we will refer to optical depth values as dimensionless). In particular it is necessary to carefully consider any systematic measurement error correlated with the brightness of the stars, such as any non-linearity of the detector or of the sky image processing methods, because those can systematically bias the fit. 

This paper is organized as follows. In Sec.~\ref{sec:fram} we describe the FRAM (F/Photometric Robotic Atmospheric Monitor) telescopes and laboratory calibration measurements necessary to properly process the acquired data. In Sec.~\ref{sec:model}  we describe the procedure to obtain a star extinction model as a function of the airmass including the molecular and aerosol components and the previous laboratory measurements. In Sec.~\ref{sec:photo} we discuss the details of the photometry (i.e. the conversion of the raw detected star signal to star brightness) and its uncertainties. In Sec.~\ref{syst} we quantify various systematic uncertainties. Finally we briefly discuss the comparison of the results with those from a Sun/Moon Photometer and possible future prospects.

\section{FRAM telescopes}
\label{sec:fram}

\subsection{Purpose, design, and operation}

 The FRAM design stems from the need for atmosphere characterization in astroparticle-physics experiments, where ultra-high-energy cosmic rays or very-high-energy gamma photons are detected indirectly by collecting the faint (fluorescence or Cherenkov) light created during the passage through the atmosphere of the extensive air showers (cascades) started by those particles. The light from these showers can travel from a few kilometers up to tens of kilometers through the atmosphere before reaching a detector. The amount of light detected can thus be significantly impacted by the transparency of the intervening atmosphere, hence the need for accurate characterization of atmospheric extinction \citep{gencta, genauger}.

There are many options in choosing the physical setup for a wide-field aerosol measurement, depending on further applications of the instrument and the availability of complementary atmospheric data. In this paper we present results based on the data from the three prototypes of the FRAM telescopes proposed as a part of the atmospheric monitoring system \citep{fram_cta} for the future Cherenkov Telescope Array (CTA) gamma-ray observatory \citep{cta}. Two FRAM prototypes \citep{prototype} have been installed at the future CTA South (CTA-S) site in the Atacama desert in Chile near the European Southern Observatory site Cerro Paranal and one prototype at the future CTA North (CTA-N) site at Observatorio Roque de los Muchachos in La Palma, Canary Islands, Spain. The data used in this paper has been taken by the three prototypes during 2017--2019 as a part of the preliminary CTA site characterization campaign \citep{sitechar}.

The main requirement for the CTA FRAM \citep{fram_cta} is to detect changes in the transparency across the whole field of view of the Imaging Atmospheric Cherenkov Telescopes (IACTs) installed at CTA site in a single image, thus necessitating a $15\times15$ degree field of view to include a significant margin \citep{ctacalib}. The CTA FRAM design is based on modified version the of the previous FRAM telescope \citep{fram,augerframnew} installed at the Pierre Auger Observatory \citep{auger} for several applications \citep{malaga1,malaga2}. The current version of the Auger FRAM wide-field imager has a field of view of $7\times7$ degrees in order to cover the portion of a cosmic-ray shower track visible to the Fluorescence Detector of the Pierre Auger Observatory without gaps in a reasonable amount of time, as its main purpose is the detection of clouds along such tracks \citep{rapid}. Both the Auger and CTA FRAMs are thus wide-field instruments that can perform continuous scans across large spans of airmass in a few minutes using 30-second exposures -- this has the advantage that any clouds in such a scan are readily visible and thus cloud-free scans suitable for aerosol measurements may be selected based on FRAM data alone. It is however in principle foreseeable to use an imaging system with a smaller field of view, taking images in a discrete pattern across different airmasses, as long as a reliable source of external cloud veto is available, such as a high-resolution all-sky camera \citep{asc}. The balance between the benefits of the possibly larger aperture of such system against the drawbacks due to the sparse sampling and the fewer bright stars covered (for which the most reliable catalogue data exist) would have to be weighed when designing such a system. This is, however, outside of the scope of this paper, as the parameters of the FRAMs are set by external requirements.

Specifically, the CTA FRAMs use a 135\,mm f/2 (7\,cm diameter) Zeiss telephoto lens and a Moravian Instruments G4-16000 monochrome CCD camera with a $36\times36$\,mm 16-megapixel chip to obtain the required $15\times15$ degree field of view. The cameras are equipped with photometric Johnson-Cousins BVRI filters, but only the B-filter data are used in this work. The B filter has its maximum transmission around 420\,nm that is located at the short-wavelength end of the spectral transmissivity of the lens used (see Sec.~\ref{sec:spectral_responce} and Fig.~\ref{fig:dusan}), and avoids almost all molecular absorption bands. Therefore the molecular contribution to the VOD using this filter is almost completely described by Rayleigh scattering and thus depends only on the integral air column above the telescope, which is easily available from global weather models and has little variation with time. It would be possible to use a different selection of narrowband filters so to avoid the atmospheric absorption bands altogether -- or, for any other reason simplifying the analysis. However the use of standard Johhson-Cousins filters facilitates the comparison with various astronomical catalogues and also allows further use of the data for astronomical purposes -- even though the catalogue we use, Tycho2, incidentally uses a slightly different bandpass and thus some conversion is still necessary.

The CTA FRAM prototypes, deployed for the purposes of atmosphere characterisation on the site, take altitude scans from horizon to zenith in 7 images with 30 seconds of exposure per image most of the nights; some of the CTA-N data has been taken with a higher density of images close to the horizon and a gap in small zenith angles for some time due to a software problem, but this does not seem to induce any systematic effects in the results.

For all images the exposure start time is synchronized to UTC time scale using the NTP protocol to an accuracy better than 1 second and all images in a scan are taken with precisely the same exposure length. Due to a software issue, for some time we had an occasional problem of the telescope mount still slewing during the start of an exposure. This may cause visible trails from bright stars in the image, but those are sometimes hard to detect visually, and virtually impossible to detect algorithmically due to the variations in sky background and noise. The main effect of this problem was that the actual star images then corresponded to less than 30 seconds of exposure (as some of the light was lost in the faint trail during the movement); if this happened to the image taken at the lowest altitude, the fitted VAOD would be biased by about 0.1, causing visible outliers in the time series of measured VAOD. Affected images can be identified from FITS metadata stored in the image files, specifically from the difference of requested and reported telescope coordinates at the start of the exposure and all affected scans have been filtered out from the data. This experience shows the importance of registering as much technical information as possible during the measurements.

\subsection{Spectral response}
\label{sec:spectral_responce}

To model the atmospheric extinction as a function of airmass for a star observed with a telescope, it is necessary to know the spectral response of the system $R(\lambda)$, which is the normalized probability density of detecting a photon of a given wavelength (as the camera signal is proportional to the number of photons, not, for example, energy density). The properties of each element of FRAM's optics (lens, optical BVRI filters, a protective covering UV filter and the CCD camera) were measured in a laboratory. The spectral transmittance of the BVRI and protective filters was measured using a Perkin Elmer Lambda 850 spectrophotometer for different incidence angles. The results show that incidence angle has negligible impact on spectral transmittance. As the photon incidence angles are relatively small in our telescope FoV we could neglect the angular dependence of the Fresnel losses. The spectral transmittance of the telephoto lens varies with the distance from the optical axis due to different thickness of the lens  elements, but we verified that its spectral dependence is negligible. The measurement setup uses a wide wavelength range light source (LOT ARC light source with 350\,W Xe Hg bulb) to cover the whole transmittance window of the lens. The light source is connected via optical fiber to a collimator illuminating a small region of the lens aperture. After passing through the lens, the light was collected at the focal point by the integrating sphere Labsphere 3P-GPS-053-SL and delivered via optical fiber to the Avantes AvaSpec Dual-channel Fiber Optic Spectrometer. All work was done in a dark laboratory to exclude stray light.  A reference spectrum from the lamp was first obtained without the lens, then the lens was measured both centered (aligned with the optical axis) and at several offsets perpendicular to the optical axis to scan the transmittance for different radial distances. The lens was tested with the iris fully open at f/$2.0$ as during the standard operation of FRAM. After subtracting the spectrometer dark current, the lens spectral transmission could be calculated from the ratio of two spectra measured with and without the lens. The measurement setup includes an optical filter for the filtration of secondary difraction orders. 

\begin{figure}[tbh]
\begin{centering}
\includegraphics[width=\textwidth]{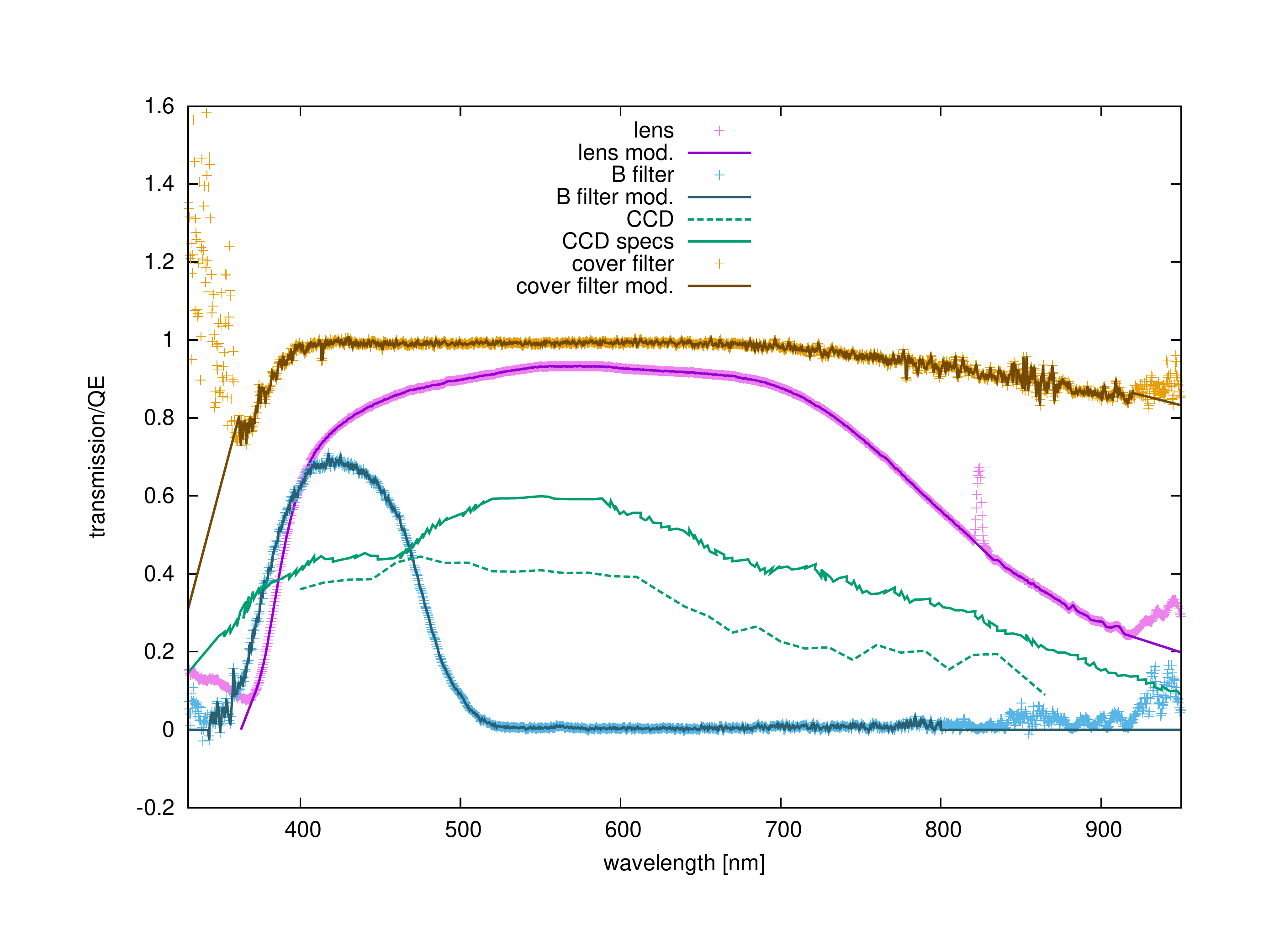}
\par\end{centering}
\caption{Measured spectral transmittance for the optical elements and quantum efficiency of the CCD sensor used in the CTA-N FRAM setup (points and dashed line). The modified and manufacturer supplied forms used for the calculation of $R(\lambda)$ are shown as solid lines. Uncertainties of the individual measurements are not shown.}
\label{fig:dusan}
\end{figure}

The transmittance measurements provide reliable results only in the interval from 365--900\,nm, even though the working range of the Avantes spectrometer is declared as 170--1000\,nm by the manufacturer. Outside of this range, high signal fluctuations are observed because the optical collimator used was made from BK7 glass which severely restricted the useful spectral range of the lamp and thus the results were dominated by the noise in the reference measurement -- we thus use smooth extrapolations of the measured data in outside of the reliable range, see Fig.~\ref{fig:dusan}. The B filter is also made from BK7 glass and thus we do not expect it to exhibit significant transmission in the range where the signal is limited by the collimator. The prominent peak at 827\,nm in the lens transmissivity curve is caused by an emission line of the light source which was likely saturated in the reference measurement, but it has no effect on the combined results for $R(\lambda)$  as the transmissivity of the B filter there is minimal.

The relative quantum efficiency of the camera CCD chip was measured using the same broad band light source connected using an optical fiber to a monochromator that allows to select the wavelength with specified full width at half maximum (FWHM). The output light of the monochromator is connected to a variable attenuator and then divided equally into two (the splitting ratio is wavelength independent). One of the fiber ends is coupled with the Thorlab PM100USB calibrated photodiode (the accuracy specified by the producer is better than $1\%$). The second fiber end is placed in front of the CCD chip to illuminate a small part of its surface. Once the wavelength is selected and the intensity is set using the attenuator to the nominal value (1000 pW)the exposure can be started. The dark frame is also acquired with open shutter and attenuator transmission set to $0T$. The precision of the intensity of the light on the output of the fiber is given by the accuracy of the calibrated photodiode, the precision of attenuator and fluctuation of the light source (monitored via the PM100USB control unit). The total intensity of the light illuminating the monitoring photodiode is ($1000 \pm 50$)\,pW. 

The relative spectral quantum efficiency of the cameras was measured only in a limited range and with a limited sampling in wavelength; we use the manufacturer specifications instead, with the laboratory QE measurement serving as a validation (at least in the relevant bandpass of the B filter). The measured QE clearly differs from the specifications, likely due to the shortcomings of the measurement setup, in particular the laborious manual adjustment of the intensity for each wavelength. In any case, the QE of the camera varies slowly enough within the B filter transmission region so that its effect on the spectral response is overshadowed by that of the other optical components and the difference between using the measured values or the specifications on the results of aerosol measurements is negligible.

The overall $R(\lambda)$ is obtained simply by the multiplication of the contributions of individual components. The effects of the uncertainties in the determination of $R(\lambda)$ on the aerosol measurements are explored in detail in Sec.~\ref{syst}.

\subsection{Detector non-linearity}

\begin{figure}[tbh]
\begin{centering}
{\includegraphics[width=\textwidth]{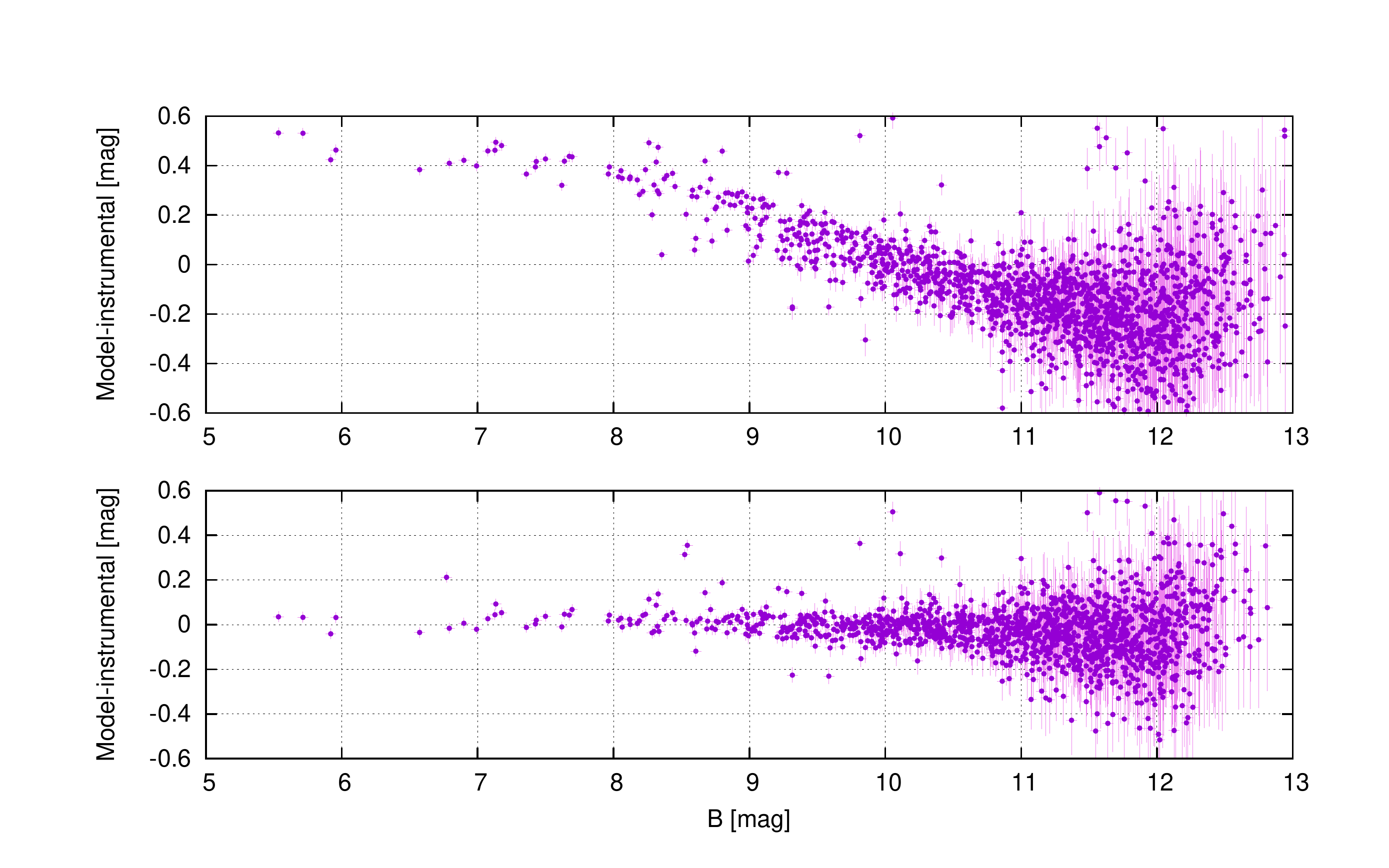}}
\par\end{centering}
\caption{Photometric residuals after fitting the instrumental star magnitudes measured on a single CCD frame with low mean intensity level with a model including Tycho2 \citep{tycho2} catalogue magnitudes, color term corrections to account for a difference in photometric system, and low-order spatial polynomial to fix residual uncorrected vignetting. Upper panel -- original data residuals, lower panel -- the residuals after additional correction for detector response non-linearity.
Outliers correspond mostly to blended stars where the aperture photometry we use performs badly, as well as to intrinsic photometric uncertainty of the Tycho2 catalogue and a difference of photometric systems.
\label{fig_photometry}}
\end{figure}

The dark- and bias-subtracted CCD signal is often assumed to be a linear function of the incoming light, at least for values well below the maximum dynamic range of the camera. However initial photometric analysis of individual frames acquired by several FRAM telescopes, as well as the spurious dependence of measured VAOD on sky background \citep{atmohead18} have demonstrated the presence of an unaccounted effect in the data analysis, which looks like to be due to significant detector non-linearity, especially in a low light level regime (see Fig.~\ref{fig_photometry}). Therefore we developed a dedicated measurement routine, applicable both in the laboratory \citep{karpov_nonlin_2018} and on the remote telescope location, in order to study and characterize the detector response linearity.

\begin{figure}[tbh]
\begin{centering}{\includegraphics[width=\textwidth]{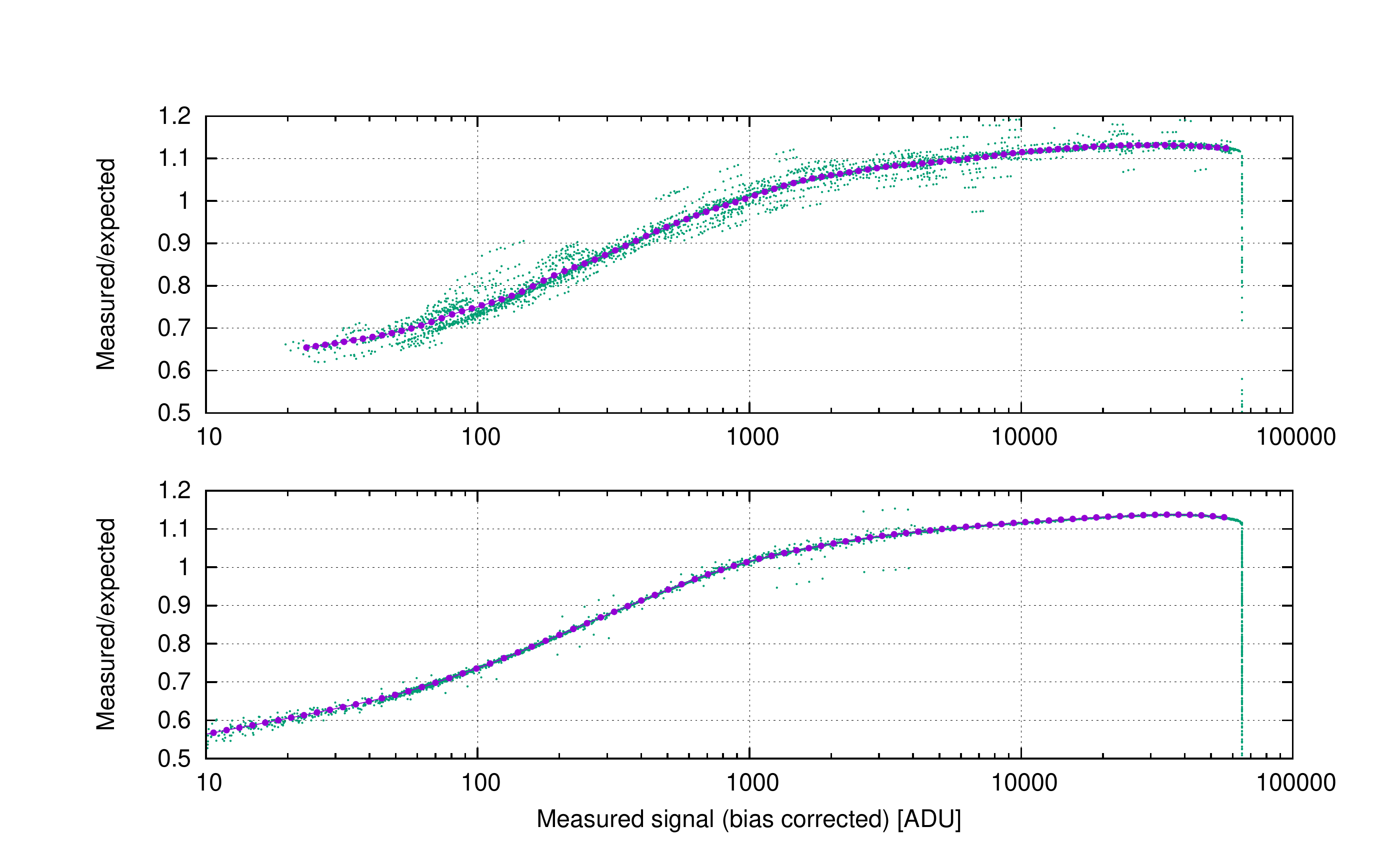}}
\par\end{centering}
		\caption{Results of non-linearity measurement for the same CCD camera (serial number 6069), first in the remote telescope location (upper panel) and after on the dedicated calibration stand in laboratory (lower panel). The saturation effects are obvious above approximately half of dynamic range ($\sim$30000\,ADUs, corresponding to $\sim$48000\,e$^-$). Non-linearity is defined as the ratio of actually measured signal to the expected one. Every point corresponds to the mean value of a signal over large (tens of thousands of pixels agreeing on the incoming light intensity to 1\%) regions of a single frame. The mean values of the non-linearity are over-plotted in logarithmically spaced bins as violet dots. The outlier points from the mean curve mostly correspond to time intervals of rapid changes of incoming light intensity, which could not be properly reconstructed from ``calibration'' frames, and are insignificant for the determination of the non-linearity.
The detector response is non-linear over the whole dynamic range.
\label{fig_nonlin_single}}
\end{figure}

\begin{figure}[tbh]
\begin{centering}
{\includegraphics[width=\textwidth]{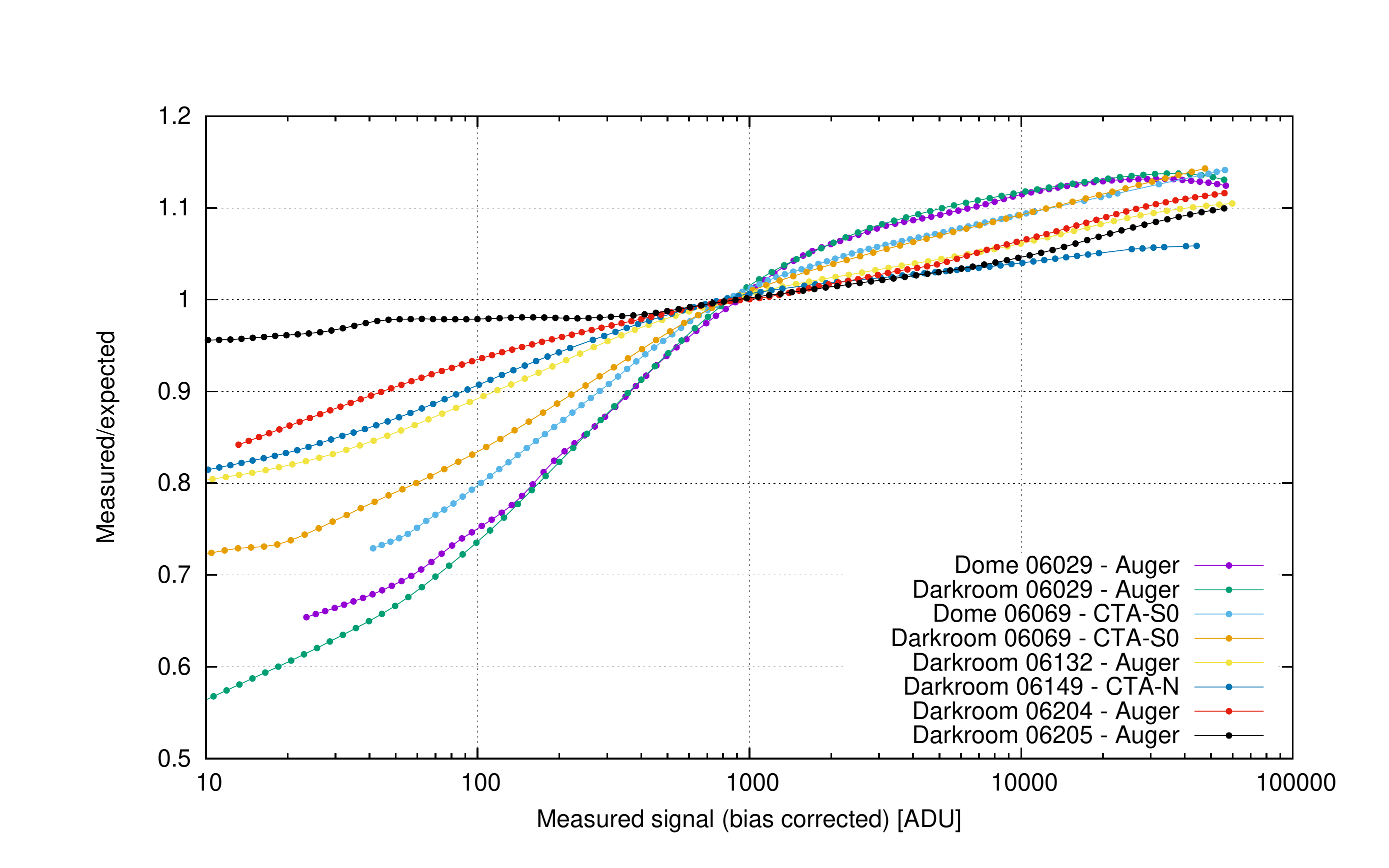}}
\par\end{centering}
\caption{Results of non-linearity measurement for a number of CCD cameras used by the FRAM telescopes installed at the future CTA sites as well as at the Pierre Auger Observatory \citep{augerframnew}. The cameras are based on the same CCD chip and were mostly measured on a dedicated calibration stand under dark room conditions. There is a progressive decrease in non-linearity with increasing serial number, which may correspond to the gradual improvements of the manufacturing process and electronics used in the cameras.
\label{fig_nonlin_all}}
\end{figure}

In order to characterize the linearity properties of G4-16000 cameras under specifically prepared experimental protocol we used a dedicated calibration stand in FZU -- Institute of Physics in Prague. The stand utilized an uncalibrated 405nm LED light source and a screen placed in front of the camera illuminated in a smoothly non-uniform way so that every acquired frame covers a range of intensities simultaneously. The measurements at the actual telescope location were organized during day time with a closed dome, using scattered light and a section of the inner surface of the dome as a non-uniformly illuminated screen. Such a setup was chosen as it was possible to implement the calibration measurements without physical access to the remote telescope site.

The acquisition protocol consisted of obtaining a number of images with different exposures and then studying the dependence of signal level on exposure time. To do so, we acquired a series of ``light'' measurements, with shutter fully open, each immediately followed by a ``dark'' measurement of the same length with closed shutter to subtract the bias level of the images. To exclude any possible bias related to changes in the light source intensity, environmental parameters and camera electronics, we randomly sampled the exposure interval from log-uniform distribution between 0.1\,s and 300\,s. Moreover, to monitor the stability of the light source, after every ``light and dark'' image pair we acquired a similar pair of ``control'' images with an exposure interval arbitrarily fixed to 10\,s. In the laboratory we repeated the acquisition sequence with different amounts of light reaching the detector. In the telescope experiment setup, the intensity of light was constantly changing due to day/night cycle. Every run contained several hundreds to thousands of images acquired over long time intervals, from several hours up to several days.

To speed up the frame read-out, we acquired only a $1024\times1024$ pixel sub-region of every CCD image frame (either in the chip center or, later, in the corner of the chip in order to contain the overscan region), which resulted in a 6 second read-out time. Then we subtracted the bias level from every ``light'' image, isolated the regions with similar illumination intensities\footnote{This procedure is done once for every continuous sequence of frames using in a single medium-intensity image on which sufficiently large (having more than 10000 pixels) sub-regions are identified so that the intensities within them do not vary by more than 1\%.}, and computed the mean signal values over these regions to be used in the analysis.

We define the detector non-linearity $N$ as the ratio between the measured signal level $I_{\rm obs}$ and the ``expected'' value directly proportional to the exposure time $T_{\rm exp}$ (corrected for shutter effects as described in \citep{karpov_nonlin_2018}),
\begin{equation}
N(I_{\rm obs}) = I_{\rm obs} / \left( F \, T_{\rm exp} \right)\ \mbox{,}
\label{eq_nonlin}
\end{equation}
where $F$ is a scaling coefficient proportional to the incoming-light flux level defined it in such a way that the mean non-linearity is equal to unity over the interval of intensities between 300 and 3000\,ADU (analog-to-digital units).
In all our setups the intensity of illumination varied slowly over time, either due to the variations of the light source intensity with temperature in the calibration-stand setup, or due to changes of the in-dome illumination related to the motion of the Sun and clouds in on-telescope setup. Therefore, we reconstructed the per-frame values of $F$ using the recorded ``control'' images all having constant exposure of 10 seconds as
\begin{equation}
F \propto I_{\rm obs,10} / N (I_{\rm obs,10}),
\label{eq_calib_time}
\end{equation}
where $I_{\rm obs,10}$ may be interpolated at any given moment, assuming smooth variations of incoming light, and the $N$ term accommodates for the fact that ``calibration'' frames are also affected by the detector response non-linearity.

Finally, we may iteratively solve Eq.~(\ref{eq_nonlin}) together with Eq.~(\ref{eq_calib_time}) to reconstruct the detector non-linearity as a function of measured signal level. Fig.~\ref{fig_nonlin_single} shows an example result of a single detector characterization, first done directly on a remote telescope site, and, later after the return of the camera back to Prague, in a dark room environment using CCD overscan regions to better mitigate the variations of a bias level. It is clear that both measurements provide the same non-linearity curve, despite the difference in experiment conditions and illumination intensities. The camera response is significantly non-linear, with measured signal level directly proportional to the exposure time (manifesting as a horizontal segment on the plot) only in a small part of the dynamical range. Instead, the non-linearity curve shows a characteristic shape consisting of several (two or maybe even three) log-linear segments with different slopes, with non-linearity being the most significant in the low intensity region. 

The shape of non-linearity curves is similar among all cameras we tested, with a progressive improvement in those manufactured more recently, i.e.\ having higher serial numbers (see Fig.~\ref{fig_nonlin_all}). They all show a characteristic structure with a slope change at around 1000\,ADU, which may either be tabulated, or approximately fitted by separate log-linear segments. Finally, the signal on the detector may be linearized by dividing the observed bias and dark subtracted value with $N$ as:
\begin{equation}
I_{\rm linear} = I_{\rm obs} / N (I_{\rm obs}).
\label{nlc}
\end{equation}

The lower panel of Fig.~\ref{fig_photometry} shows the effect of a linearization of the data on photometric residuals.

\subsection{Data processing}

The raw CCD frame is first corrected by subtracting dark counts (due to the dark current in CCD chip) and bias level (due to A/D electronics) signals. For most purposes, subtracting a dark frame taken at the CCD temperature (or rather, to avoid introducing noise, a median of many dark frames) can satisfactorily remove both dark counts and bias, leaving only a small residual offset due to the bias drift with electronics temperature and history. However, while the dark signal depends only on the (stabilized) temperature of the CCD chip, the bias signal depends on the temperature of the camera electronics, which are outside of the cold chamber and therefore the bias signal varies both with environmental conditions and the history of its load. The bias frame is determined either using short dark exposures (the bias does not vary with exposure time, while the dark current for a short exposure is effectively zero) or just by scaling a normal dark frame according to the current camera electronics temperature -- as the bias component is so much larger than the dark, these methods give equivalent results (see Sec.~\ref{syst}). The non-linearity correction (NLC) described above turns out to be extremely sensitive to any residual offset and so the precise bias for each image must be  subtracted. Originally, this was done using the observed correlation between the bias level and the electronics temperature, determined over a very large number of dark frames at constant CCD temperature acquired over many months of operation (the dark current at the nominal operating temperature $-20$\,$^{\circ}$C is very small compared to the bias variations for these cameras). Later, the firmware of all our cameras was modified to read out the so-called overscan region: the area of the CCD chip outside of the nominal imaging area. This area is physically masked from light illumination and thus provides information on the dark+bias signal simultaneously to any camera acquisition. The average (median) of the overscan region is then subtracted to the relative CCD acquisition frame thus correcting for any drift of the bias with temperature.

With the bias and dark components subtracted, each image is linearized using Eq.~(\ref{nlc}) and then divided by a flat-fielding image produced from the combination of many images taken during dusk/dawn in the direction of the sky with the lowest expected brightness gradient to obtain uniform illumination -- this is a standard procedure to correct for vignetting (which is significant with the lenses used) and variations in detection efficiency across the CCD chip.

Astrometric calibration of the images is then performed using the local installation of Astrometry.net \citep{astrometry.net} code, which provides a WCS solution typically accurate to better than 1 pixel over the most of the field of view. The point sources are detected in the calibrated image using the SExtractor software package  \citep{sextractor}, and then identified with stars from the Tycho2 catalogue \citep{tycho2} if their predicted position on the image lies within a short distance (typically 8 pixels) from the position of the detected object centroid. Any pair of sources detected within a close distance (set at 20 pixels) from each other are rejected to avoid mutual influence during photometry (blending). Only stars within 1840 pixels from the image center are taken into account, corresponding to 63\% of the total field of view, as beyond this limit, the PSF of the stars deteriorates quickly and depends strongly on the momentary focusing of the lens. Also rejected are stars close to saturation, where strong non-linear effects appear both due to saturation itself and due to the anti-blooming gate of the CCD used. The procedure of evaluating the brightness of individual stars from an image in not unequivocal regarding in particular the choice of the software for aperture photometry (we use both SExtractor itself and IRAF \citep{iraf,iraf2} apphot package) and the radius of the photometric aperture in pixels -- the appropriate choice as well as the associated uncertainties are discussed in Sec.~\ref{photc}.

\section{Data model}
\label{sec:model}

\subsection{Model fitting}

In general, the measured (instrumental) star brightness $m_\mathrm{inst}$ is related to the catalogue value $m_\mathrm{cat,B}$ by the equation \citep{atmohead16}
\begin{equation}
\frac{m_\mathrm{inst}}{M}=m_\mathrm{cat,B}+Z_i+g(A_\mathrm{M},A_\mathrm{A},A_\mathrm{O},k_i,B-V)+f(B-V,x,y) \label{model},
\end{equation}
where $f(...)$ describes corrections due to the different response of the system to stars of different color index (as expressed by difference between the Tycho $B$ and $V$ magnitudes\footnote{In the following, $B-V$ will always refer to the difference between the magnitudes of a star in B and V filters \emph{in the Tycho photometric system}, i.e. B$_{\rm T}$ and V$_{\rm T}$ \citep{hipparcos_passbands}.}) and in different parts of the field of view, and $g(...)$ is a model of extinction as a function of the airmass, the color index of the star and the actual value of the VAOD (described in detail in the following section); $Z_i$ is the momentary absolute calibration constant of the system (so-called zeropoint) and $M$ is a correction constant for any possible non-linearity introduced in the data during the photometric process. Note that $Z_i$ also automatically absorbs any effect due to the uncertainty of the overall normalization of the catalog fluxes as discussed for example in \citep{HST}. 

The atmosphere is composed of several components with different vertical profiles -- for each of these components, the airmass $A$ is a different function of the altitude of the star above the horizon.  In our case, we consider $g(...)$ as a function of three different airmass values:  $A_\mathrm{M}$ describes the molecular atmosphere according to \citep{airmassM}; $A_\mathrm{A}$ describes the tropospheric aerosols using the formula for water vapor from \citep{airmassA}, as the relevant vertical distributions are similar according to \citep{airmassthesis}; $A_\mathrm{O}$ is the airmass for ozone accroding to \citep{airmassO} used in a similar way as an approximation for the vertical distribution of stratospheric aerosols. The formulae for water vapor and ozone should in principle be also used for the respective components of the molecular atmosphere, but in the passband given by our B filter, the total contribution of water vapor and ozone to molecular extinction is only 0.001 and thus the effect of their different airmass formulae is negligible. Each airmass value is calculated from the average altitude of the star above the horizon (corrected for refraction) during the exposure -- a more precisely correct way would be to take the weighted mean of airmasses at several points during the exposure with the weight being given by the transmissivity of the atmosphere at that airmass, but the difference with respect to a simple average of altitude is less than 0.0003 in airmass.

The correction for the $B-V$ color term in $f(...)$ is taken as a third order polynomial, but only stars with $ B-V<0.8$ are used; for stars with larger color index, parametrization proves difficult. As the FRAM spectral response does not faithfully reproduce the Johnson system, the resulting correction differs from the nominal Tycho2 conversion formula \citep{tycho2} $B_\mathrm{Johson}=B_\mathrm{T}-0.24(B_\mathrm{T}-V_\mathrm{T})$ by up to 0.015 mag for $B-V\in[-0.1;0.8]$ and deviates more for even bluer stars, which are however rare in the sky.

In principle, the spatial dependence of $f$ can reflect residual imperfections of normal flat-fielding correction but an additional spatial dependence is introduced by the photometric aperture method, in fact the fraction of the total star light captured by the photometric aperture is strongly dependent on the star's PSF which, for wide-field setups, is in turn strongly dependent on the star's position on the chip. Ideally, the spatial dependence would be modeled by a simple function of the star distance from frame center, but despite our best mechanical efforts, each hardware setup shows a deformation of the pattern of the residuals in some different direction, likely due to a combination of non-orthogonality of the lens mount and imperfect adjustment of the orthogonality of the CCD sensor to the optical axis. To find a ``photometric flat-field'' correction map we first take a set of scans and process it without any further position-dependent correction and then calculate the residuals of the fit for all stars as a function of the position on the chip and smear this data with a 100-pixel Gaussian filter to produce a correction map which is then used as an additional flat-field on every processed image (technically it is then no longer part of $f$ during the fit).

\begin{figure}[tbh]
\begin{centering}
\includegraphics[width=0.60\textwidth]{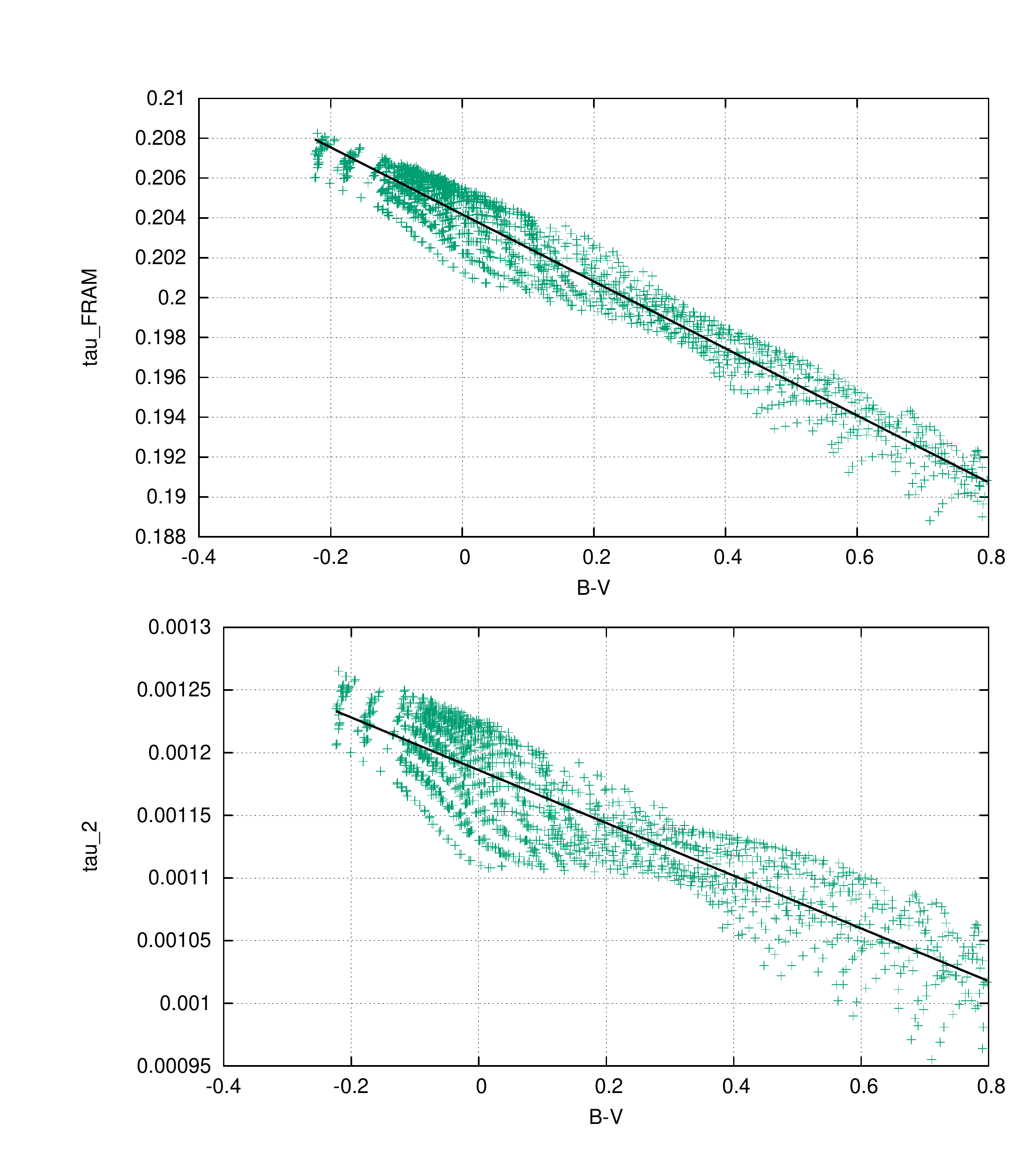}
\par\end{centering}
\caption{The coefficients $\tau_\mathrm{FRAM}$ and $\tau_2$ as obtained by fitting a star extinction for individual synthetic stellar spectra as a function of the spectra $B-V$ color index. Continuous lines are the linear functions used to approximate these two coefficients as a function of the star color index.}
\label{fig:ray}
\end{figure}

\subsection{Molecular extinction}

For light of a given wavelength, the attenuation in magnitudes as a function of airmass $A$ is simply
\begin{equation}
\Delta m = 2.5 \log_{10} \frac{I_\mathrm{ground}}{I_\mathrm{space}} = 2.5 \log_{10}(\exp(\tau(\lambda))\,A) = \frac{2.5\,\tau(\lambda)\,A}{\ln10} \approx 1.086\,\tau(\lambda)\,A = k(\lambda)\,A
\end{equation}
where $k=1.086 \tau$, is the extinction coefficient in magnitudes. For molecular extinction, we obtain $\tau(\lambda)$ using the MODTRAN software package \cite{modtran} based on a typical vertical profile of the density of the atmosphere for the FRAM locations from the ECMWF ERA5 data\footnote{https://www.ecmwf.int/en/forecasts/datasets/reanalysis-datasets/era5}. We are in principle able to obtain the profile specifically for the exact time of each observation (within the time resolution of the atmospheric data), but the difference with respect to using the mean profile is less than 0.001 in VOD for CTA-N and less than 0.0003 for CTA-S, based on a test calculation using 60 different profiles across one year. For CTA-S, this variation may be underestimated in the global models due to the lack of regular vertical profile measurements in the area and is subject to current research  \citep{Julio}.

In order to study the dependence of the molecular extinction as a function of the airmass and spectral color we utilize a library of theoretical stellar spectra \citep{jakub1}, from which we extract the normalised spectral density  $F(\lambda)$ for each stellar model, and using the FRAM spectral response as obtained in Sec.~\ref{sec:spectral_responce} for $R(\lambda)$, we calculate:
\begin{equation}
\frac{I_\mathrm{ground}}{I_\mathrm{space}}=\int F(\lambda)\,R(\lambda)\,\exp{(-\tau(\lambda)\,A)}\,\mathrm{d}\lambda \approx\exp(-\tau_\mathrm{FRAM}\,A+\tau_2\,A^2), \label{extint}
\end{equation}
where the other terms in the Taylor expansion are negligible (amounting to difference less than 0.001 optical depths from the exact value for most of the spectra), while the quadratic term is important as a linear approximation would cause a difference of up to 0.03 optical depths at the $A=4$ for some spectra. Both $\tau_\mathrm{FRAM}$ and $\tau_2$ are to be understood as functions of the spectral properties of the star. For each stellar model we also compute the color index $B-V$ (using the conversions from \citep{jakub2}), which is a spectral parameter available from the Tycho2 catalogue for every star. For the region $B-V<0.8$ we find that both $\tau_\mathrm{FRAM}$ and $\tau_2$ can be well parametrized as linear functions of $B-V$ (see Fig.~\ref{fig:ray}), so that 
\begin{equation}
\tau_\mathrm{FRAM}  = k_\mathrm{M}+k_\mathrm{C,M}(B-V) 
\end{equation}
\begin{equation}
\tau_2  = k_{2,\mathrm{M}}+k_{2\mathrm{C,M}}(B-V),
\end{equation}
(with $k_\mathrm{C,M}$ and $k_{2\mathrm{C,M}}$ negative). Thus the model for molecular extinction in magnitudes as a function of airmass and color index is
\begin{equation}
g_\mathrm{M}(A_\mathrm{M},B-V) = k_\mathrm{M}\,A_\mathrm{M}+k_\mathrm{C,M}\,A_\mathrm{M}(B-V)-k_{2,\mathrm{M}}\,A_\mathrm{M}^2-k_{2\mathrm{C,M}}\,A_\mathrm{M}^2\,(B-V) \label{molmodel}
\end{equation}
 (the $k_{2\mathrm{C,M}}$ term being a small correction with an effect on VAOD estimation of the order of 0.001). We checked whether a more general function would improve over this parametrization, but did not find significant improvement as most of the remaining spread is due to other properties of the stars not reflected by their color index (see Fig.~\ref{fig:ray}). This procedure assumes that all stars from the stellar spectra occur with the same frequency, which is likely not true; moreover the relative contribution of stars with different properties may vary across the sky -- the impact of this on the uncertainty of the VAOD measurement is explored in Sec.~\ref{syst}. For the analysis of data in the B filter, where the contribution of absorption is small and the molecular extinction is dominated by the Rayleigh scattering which is directly proportional to the surface atmospheric pressure, a linear correction accounting for the variations in the pressure can be added where pressure data are available.

\subsection{Aerosol extinction}

The wavelength dependence of optical depth for general aerosols is in general assumed to be expressed by the form
\begin{equation}
\tau(\lambda)=\tau_0(\lambda/\lambda_0)^{-\alpha}\label{angst}
\end{equation}
with $\alpha\in[0;2]$ being the \AA ngstr\"om exponent and $\lambda_0$ is a reference wavelength. In our case, we put $\lambda_0=440$\,nm which is the mean wavelength of the light detected by FRAM for a star of $B-V=0$ when $\alpha=1$ after subtracting molecular extinction -- this choice minimises the uncertainty of $\tau_0$ for the given experimental setup. Repeating the integration of Eq.~(\ref{extint}) using Eq.~(\ref{angst}) for $\tau_0\in[0;0.2]$ (the typical range of conditions we find at the CTA sites) and $\alpha\in[0;2]$, we find that the tropospheric aerosol extinction is well described with the following model (with a single free parameter $k_\mathrm{A}$)
\begin{equation}
g_\mathrm{A}(A_\mathrm{A},B-V,k_\mathrm{A}) = (1+0.011\alpha)k_\mathrm{A}\,A_\mathrm{A}-0.023\,\alpha\,k_\mathrm{A}\,A_\mathrm{A}\,(B-V)-0.0022\,\alpha^2\,k_\mathrm{A}^2\,A_\mathrm{A}^2 \label{aermodel}.
\end{equation}

This indicates that in the case of unknown \AA ngstr\"om exponent $\alpha\in[0;2]$, assuming $\alpha=1$ leads to a relative uncertainty of the order of 1\% in $k_\mathrm{A}$, which is small in absolute value (we test this estimate against data in Sec.~\ref{syst}). The quadratic term in this case is negligible for all but the largest aerosol loads, but must be kept to measure those values precisely. 

Additionally to the tropospheric aerosols, there is, at any moment, also some amount of aerosols present in the stratosphere. For those stratospheric aerosols, we refer to the data from \citep{strato} which indicate that during the period covered by our measurements, their amount was relatively constant, amounting to $\tau=0.0064$ at $\lambda=420$\,nm using the measured $\alpha=1.6$. For such a small value all other corrections are negligible and the model is simply
\begin{equation}
g_\mathrm{O}(A_\mathrm{O}) = 0.0069A_\mathrm{O}.  \label{stratomodel}
\end{equation}

The aerosol extinction model is a sum of the two individual contributions and the total VAOD is thus obtained as $\tau = 0.921(k_\mathrm{A}+0.0069)$. This simple description with a constant $g_\mathrm{O}$ is valid only in periods of low global volcanic activity -- for other periods, the split between tropospheric and stratospheric aerosols may be more complicated. However, since we are measuring the combined value, the uncertainty of $g_\mathrm{O}$ only affects the quality of the description of the angular dependence of the aerosol airmass, not the absolute value of the VAOD. When tested on the data used in the next sections, we find that using $A=A_\mathrm{M}$ for all components of the extinction leads to a shift of 0.002 in the calculated VAOD, which is small, but not negligible relative to other uncertainties and thus the distinction between extinction components should be kept. The uncertainty due to the varying split between tropospheric and stratospheric aerosols will be even smaller, barring a major global volcanic event.

The complete model of atmospheric extinction, obtained as the sum of equations (\ref{molmodel}), (\ref{aermodel}) and (\ref{stratomodel}), is thus fairly complex. Previously, we have followed a simpler approach, where we not only used a single value of $A$, but also simplified the model to
\begin{equation}
g(A,B-V,k_\mathrm{A}) =(k_\mathrm{A}+k_\mathrm{G})A-k_\mathrm{C}\,A(B-V)-k_{2,\mathrm{M}}\,A^2 \label{oldmodel}.
\end{equation}
where $k_\mathrm{C}$ (which cannot be just computed for general aerosols) is fitted as a global parameter for one set of scans and the quadratic term for aerosols is neglected -- such a simplification however differs from the more detailed model by up to 0.004 in VAOD when tested on the same data, which is already a significant contribution as can be seen in section \ref{syst}.

\begin{figure}[tbh]
\begin{centering}
\includegraphics[width=\textwidth]{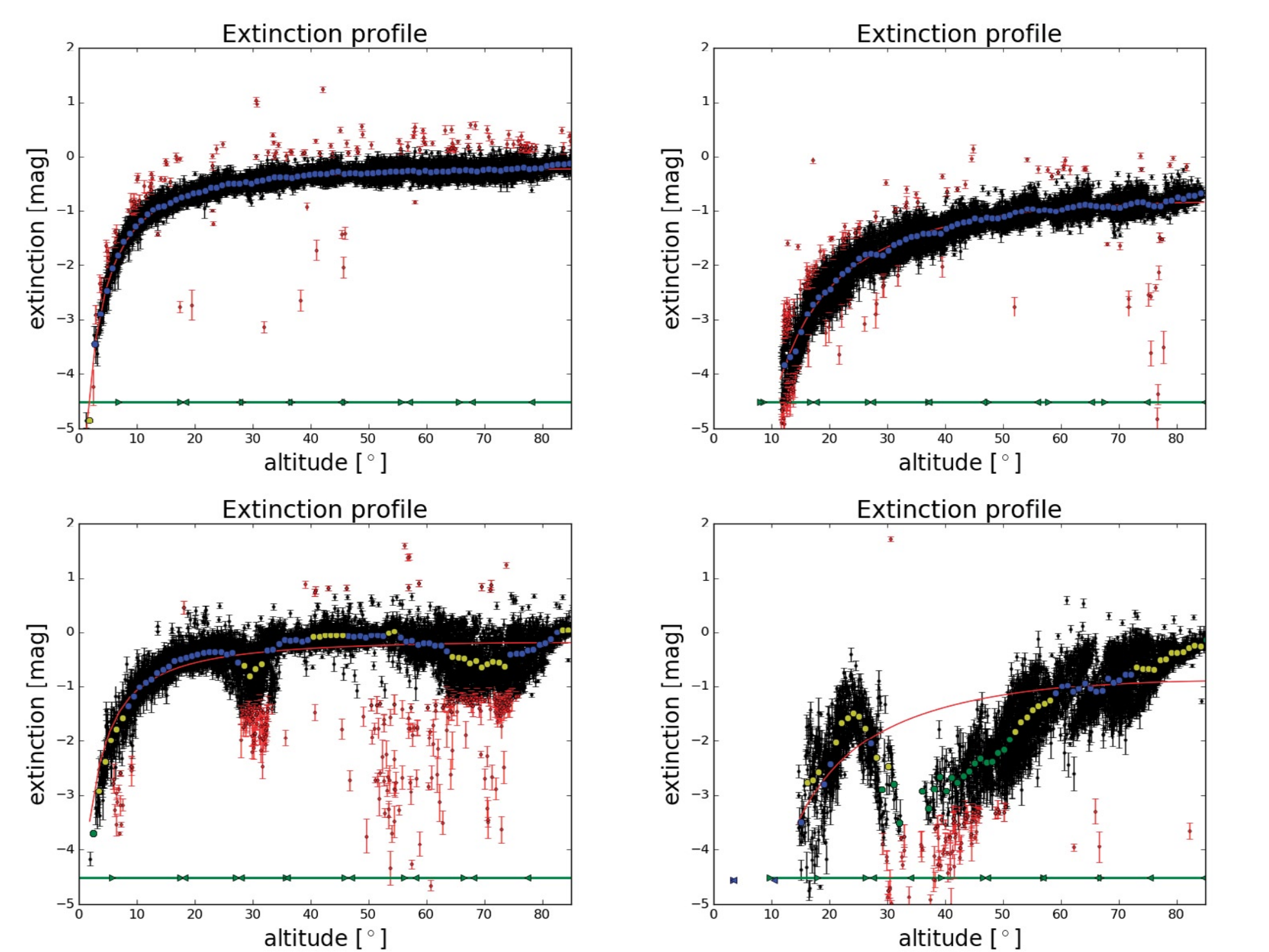}
\par\end{centering}
\caption{Examples of plots used to select clear scans for aerosol analysis from data taken at La Palma. Top left: a very clear scan (VAOD compactible with zero). Top right: a borderline acceptable scan with the best fit giving VAOD around 0.6. Bottom left: several light clouds, unacceptable scan. Bottom right: large clouds, unacceptable scan. The colors of the points indicate the mean RMS deviation of stars in an altitude bin from the preliminary fit, the green arrows show the altitude coverage of images (to show whether a gap is due to clouds or missing coverage) and stars in red were marked for rejection as outliers. Note that these are the results of the prelimiary analysis used for scan selection and the spread of the values for individual stars is larger than in the final fit used for the determination of VAOD.} 
\label{fig:ww}
\end{figure}

\subsection{Selection of clear scans}

Any scans affected by clouds must be rejected. To facilitate this process, a preliminary plot of extinction vs.\ altitude is produced for each scan by taking the correction function $f(...)$, as well as $M$ and all the parameters of $g(...)$ that can potentially vary with time, such as the optical depths for the molecular or stratospheric components, as fixed, based on a rough estimate or previously processed data (see below), and fitting a $(Z,k)$ pair to the scan. These plots (see Fig.~\ref{fig:ww} for examples) are decorated with visual clues about the mean departures from the fitted value in individual bins as well as other information, such as the altitude coverage of the individual images and identification of stars rejected by the preliminary fit as outliers, so that they can be inspected efficiently in rapid succession. Then the altitude scans deemed ``clear'' are grouped into batches of 50--100 and fitted with Eq.~(\ref{model}) simultaneously, with free parameters being pairs of $(Z,k)$ for each scan while all other parameters are considered as global variables. After the fit converges, the stars for each scan are binned in altitude and outliers above 3 standard deviations in each bin are removed and the fitting is repeated.

\section{Photometry}
\label{sec:photo}

\subsection{Error model}
\label{errm}

To compare different possible choices for the photometric extraction method, we need to associate uncertainties to the differences between the measured and catalogue brightness for the individual stars, so that they can be used as weights in the fitting procedure. To this end, we have acquired a series of several thousand images of the same region of the sky with one of the FRAM setups and extracted star fluxes from a sub-series of 700 of these for which the atmospheric conditions were exceptionally stable. The image processing was done with a photometric extraction software (IRAF with aperture radius 4) with a fixed error of 0.1 mag assigned to every star measurement. Each image was processed using the full pipeline, including near-neighbor filtering and outlier rejection. $(Z,k)$ is fitted on each image -- the values produced have large uncertainties, but the goal is testing the photometry software itself, not the determination of the VOD from this data.

\begin{figure}[tbh]
\begin{centering}
\includegraphics[width=\textwidth]{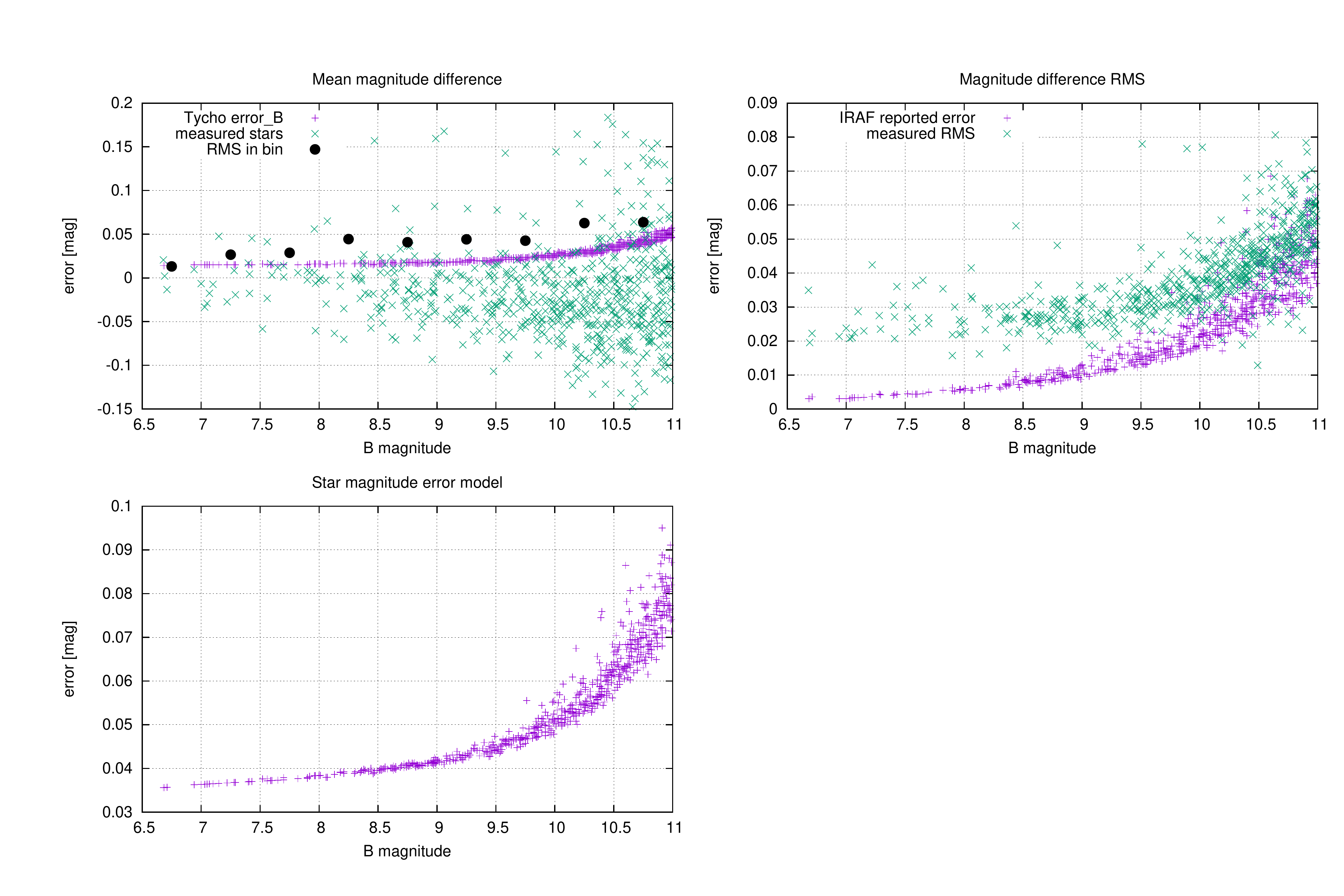}
\par\end{centering}
\caption{Results from the analysis of a long series of exposures of the same sky region. Top left: mean magnitude difference from model vs. the Tycho catalogue B magnitude with RMS (quadratic mean) for each bin shown. Top right: comparison of the actual RMS in our star magnitude measurements (green), compared to the one reported by the IRAF photometric software (violet), as a function of star B magnitude. Bottom: the resulting error model for individual star measurements in each sky image.}
\label{fig:70}
\end{figure}

For each star that is found and accepted by the pipeline in more than 10 images, the mean differences between the measured and expected star model brightness and the relative RMS are calculated; the former is then compared with the error reported by Tycho2 catalogue for the B magnitude while the latter is compared with the mean error that the IRAF software associates to the measurements of the star brightness. A systematic difference between the mean magnitude difference and the Tycho2 error is somewhat expected as there is additional uncertainty in the color correction and it can be modeled by adding a systematic error of $0.0025\,B$. The fact that the RMS is larger than the error reported by the IRAF software means that our measurements are affected by other sources of noise in addition to the uncertainty in the background estimation and the Poissionian fluctuations of the light flux that are already considered by IRAF when calculating the error on its measurement. These may be related to insufficient flat-fielding but preliminary tests show that at least a part of the effect is due to the dependence of the flux on the sub-pixel position of the star centroid (the stars' PSF are too sharply peaked near the camera center, and the CCD we use employs a microlens raster array in order to improve pixel fill factor). This additional error is modelled by adding an extra error term of 0.028 independent from the star brightness.

Overall, we can associate to each star measurement the error as (Fig.~\ref{fig:70})
\begin{equation}
\sigma^2 = \sigma_\mathrm{Tycho,N}^2+\sigma_\mathrm{phot}^2+0.028^2+(0.0025\,B)^2
\end{equation}
and similar results hold for SExtractor measured values. This formula can thus be used for estimating the errors for each star brightness that are used as weights for fitting the extinction model on the altitude scans.

\begin{figure}[tbh]
\begin{centering}
\includegraphics[width=\textwidth]{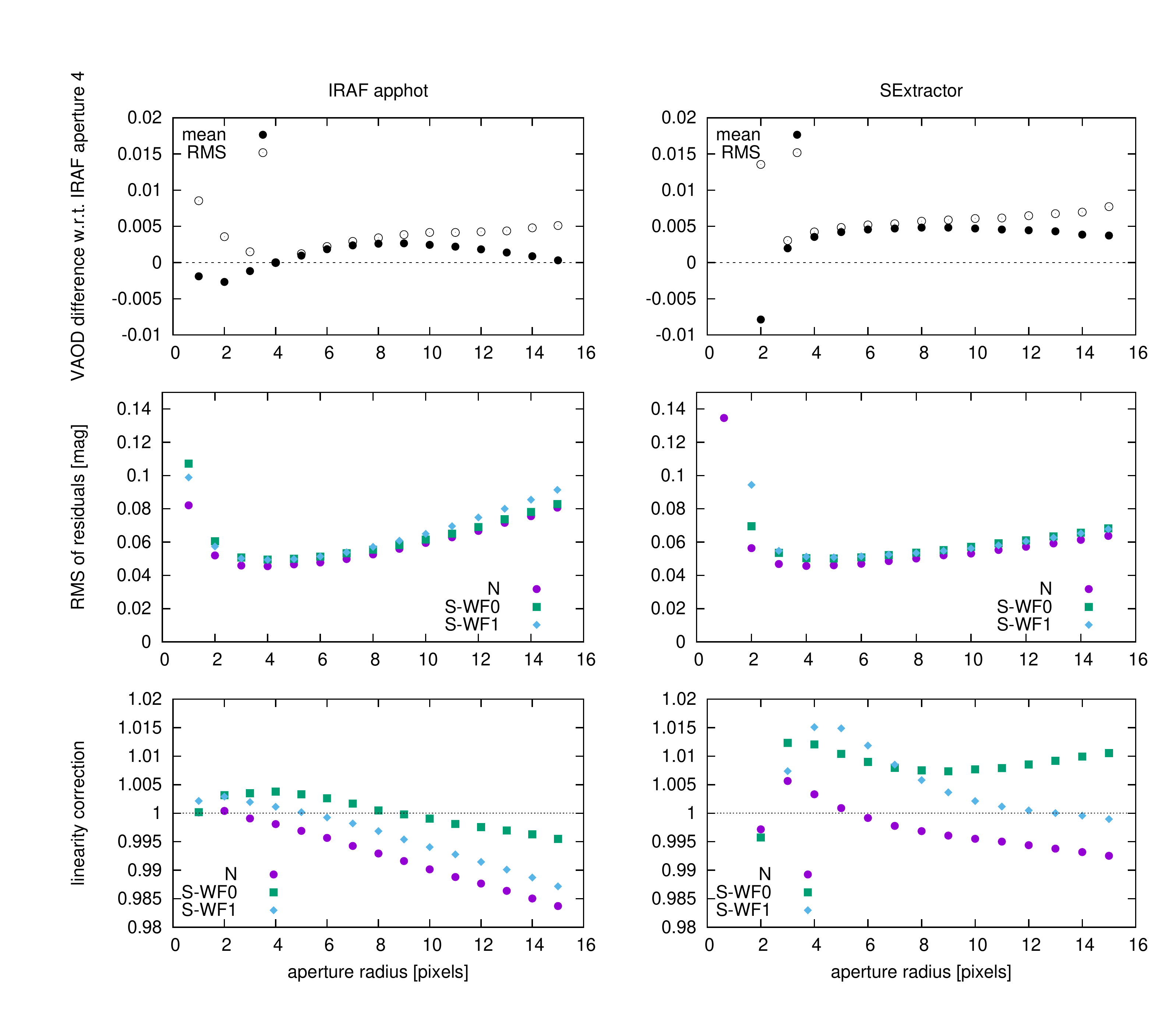}
\par\end{centering}
\caption{Comparison between two photometric software methods -- IRAF apphot vs. SExtractor as a function of the aperture radius. The plots in the first row show mean and RMS difference of the VAOD values determined from the overall set of scans, second row shows RMS of residuals for individual stars for all three FRAM setups and the last row shows the non-linearity coefficient $M$.}
\label{fig:photometry}
\end{figure}

\subsection{Choice of photometric method}
\label{photc}

In order to select the best photometric method, we compare the results obtained from the analysis of 100 randomly selected vertical scans from each of CTA-N (labeled N) and CTA-S sites (two FRAM setups labeled S-WF0 and S-WF1). The S-WF0 camera did not have the overscan readout activated, so its bias subtraction is based only on temperature parametrization. However, the CTA-N and CTA-S sites are very distinct geographically, so the actual aerosol conditions may be different. During the optimization of the photometric process, some scans appeared to be contaminated by clouds and were removed.  The results presented are thus obtained on sets of 91 (from CTA-N) and 80 and 92 scans (from each of the CTA-S setups).

\begin{figure}
\begin{centering}
\includegraphics[width=\textwidth]{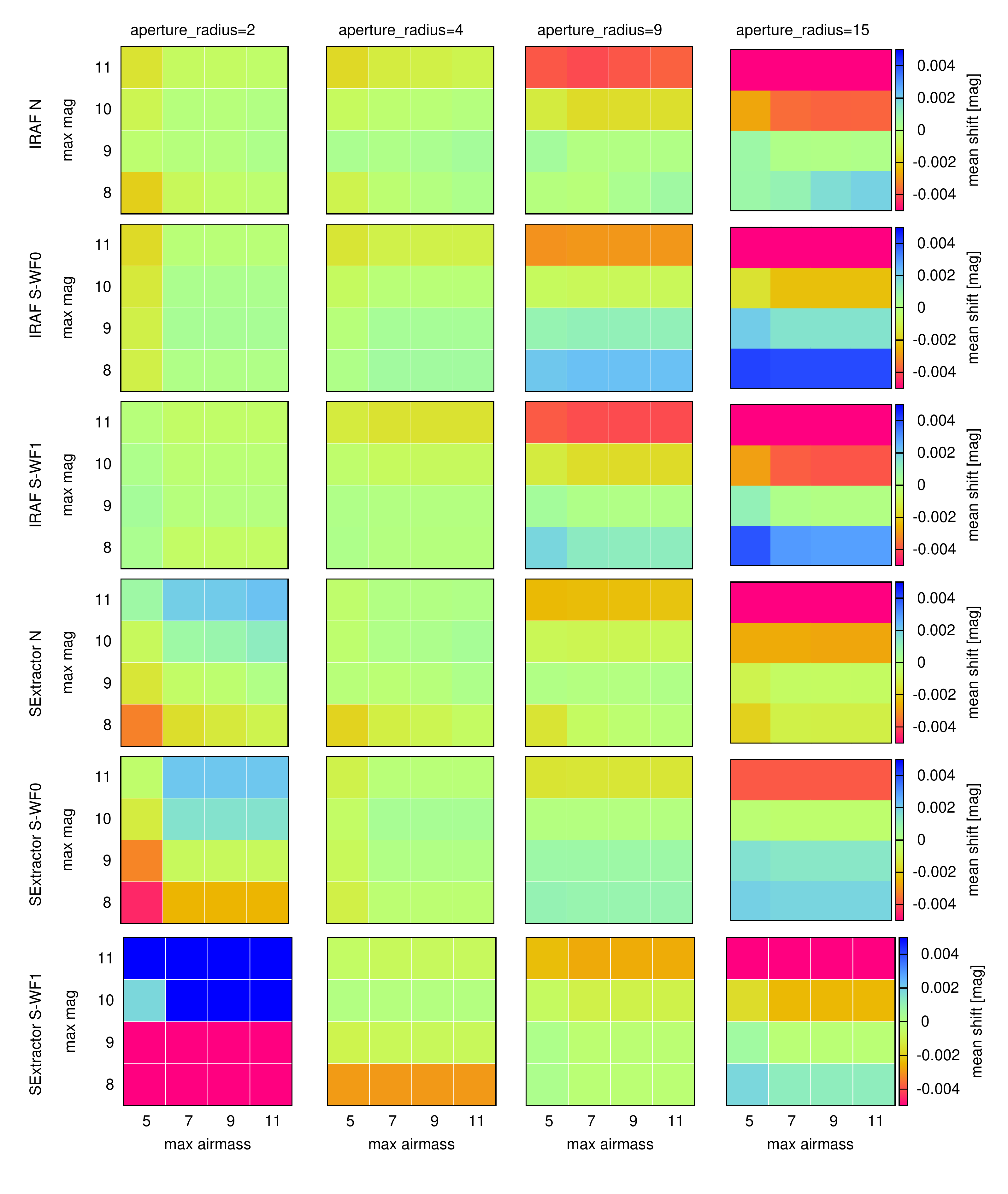}
\par\end{centering}
\caption{Mean difference in VAOD values obtained from fits with different selection cuts in star brightness and airmass (with respect to VAOD results for cuts at 9.5 magnitudes in brightness and 8 in airmass) for different photometric aperture radii (columns) and different choices of photometric methods and setup (rows).}
\label{fig:cuts}
\end{figure}

Figure \ref{fig:photometry} shows the aperture photometry results as a function of the aperture radius in pixels for both SExtractor and the IRAF apphot package. There are various differences between the two codes, most importantly in the shape of the aperture (different approximations for an intersection of a circular aperture with CCD pixel grid, especially evident for smaller radii, see e.g. \citep{aper_comp}), and the method used to extract the background -- IRAF employs local background estimation using an annulus around the star, whereas SExtractor constructs a global background model. For both codes we obtain the best agreement of the measured star magnitudes with our model for an aperture size of 3--6 pixels. We find significant differences in the non-linearity (expressed by the coefficient $M$ in Eq.~(\ref{model})) which varies between FRAM sites and it is generally more pronounced in SExtractor photometry. Disregarding 1 and 2 pixel apertures which produce unstable results (especially for SExtractor), we find that the VAOD variation as a function of aperture radius is within 0.0025, although there is a roughly 0.003 systematic difference in VAOD as measured by IRAF and SExtractor. We choose the photometry done with IRAF at aperture 4 as the reference method (all further uncertainties are evaluated with this selection).

Figure \ref{fig:cuts} shows the dependence of the fitted VAOD on the upper selection cuts imposed on the star magnitude and airmass to be accepted for the fit procedure with respect to the default configuration (used also to produce Fig.~\ref{fig:photometry}), which is $<9.5$ for brightness and $<8$ for airmass. The star magnitude cut has been chosen because the errors of the catalogue magnitudes rise sharply above this value, while the airmass cut is somewhat arbitrary. It is clear that for the aperture radius of 4 pixels, the dependence on the selection cuts is smaller than 0.0015 for all tested values of cuts especially when using IRAF software. On the other hand, for both smaller and larger apertures there may be significant shifts, mainly linked with the variation of the cut on faint stars.

\section{VAOD Systematic uncertainties}
\label{syst}

We have tested various possible sources of systematic uncertainty in the determination of the VAOD with our method using the same set of 263 scans from three different sites as in the previous section, or a subset. These include the following effects:
\begin{itemize}
\item background estimation, evaluated as the maximal difference for a set of changes in the parameters of the annulus from which the background is calculated (the thickness of the annulus set to 4, 12 and 16 pixels instead of the reference value of 8, and the inner radius of the annulus set to 6 and 12 pixels instead of the reference value of 9) as well as the change in the ``salgorithm'' IRAF setting from ``mode'' to ``centroid'',
\item CCD non-linearity correction, estimated by varying all the parameters of the fit applied to the laboratory data by 1 standard deviation,
\item bias subtraction, especially important when the overscan data are not available and thus tested on CTA-S WF0 data by varying the calibration curve between the bias signal and the electronics temperature within the range allowed by the dark frame data,
\item ``processing'' uncertainty given by the VAOD difference obtained between two processing codes which differ in the choice of reference dark frames as well as in whether they consider dark and bias signal separately,
\item the numerical ``photometric flat-field'', estimated by comparing the results obtained by creating the flat-fielding data from two different halves of the test dataset,
\item the unknown \AA ngstr\"om coefficient, quantified simply by putting $\alpha$ equal to 0 and 2 in Eq.~(\ref{aermodel}),
\item the uncertainty of the star magnitude error model, compared with the extreme case of same uncertainty for every star measurement.
\end{itemize}
These effects combined amount to a total uncertainty of less than 0.001 in the determination of the VAOD. Thus the dominant sources of uncertainty are the choice of photometry software and selection cuts on maximal airmass and magnitude of stars in the fit and the determination of the spectral response $R(\lambda)$ that affects the modeling of molecular extinction.

The uncertainty related to the choice of photometry method may be estimated as the maximal difference for IRAF settings between apertures of 3--6 pixels and SExtractor settings between 3--8 pixels, as those roughly show the minimal RMS of the model description. This uncertainty may be split into a correlated part (corresponding to the shift of the mean value of the VAOD) of 0.005 and an uncorrelated part (calculated as the RMS of the differences of individual VAOD shifts with respect to the mean) of 0.003. Similarly for the dependence on the airmass/magnitude selection cuts, the correlated part is 0.001 and the uncorrelated part is estimated as 0.003 (it is slightly larger for the smallest cuts in airmass and magnitude, but those significantly reduce the number of stars available for some scans and thus increase the statistical uncertainty).
\begin{figure}
\begin{centering}
\includegraphics[width=\textwidth]{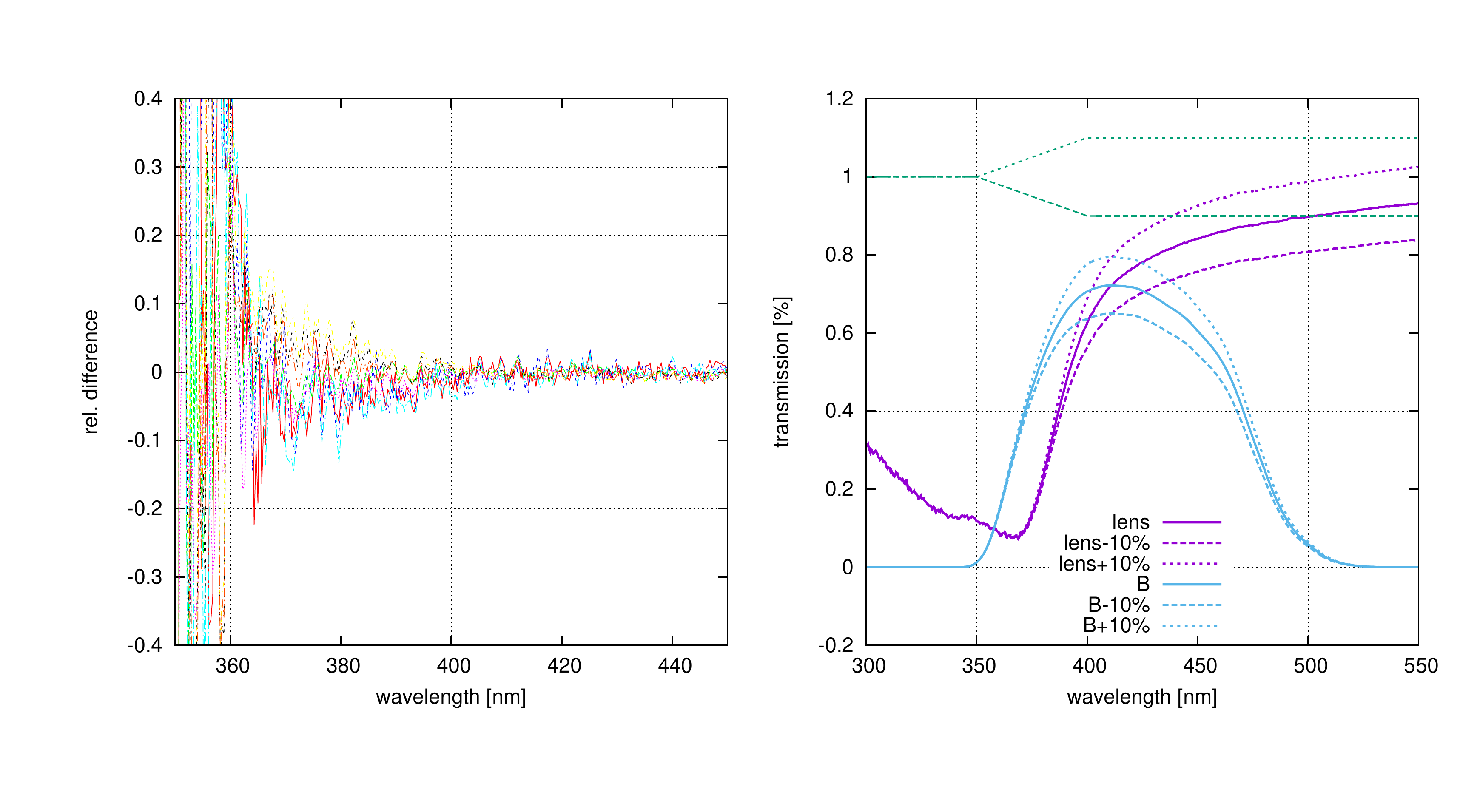}
\par\end{centering}
\caption{Left: comparison of various measurements of the transmissivity of the B filter for different integration times and light fluxes. Right: the worst case model used to put an upper limit on the uncertainty due to the determination of $R(\lambda)$.}
\label{fig:dusanerr}
\end{figure}

Determining the uncertainty of the measurement of $R(\lambda)$ is not straightforward. The key question for our purposes is the spectral dependence, not the absolute normalization of the transmittance, which has no effect on VAOD determination from the extinction model. Since in our spectral calibration method the signal measured through the optical element is immediately compared to a reference measurement, most wavelength-dependent systematics should cancel out, with the exception of non-linearity of the light detector. We have thus carried out a series of measurements (using the B filter as an example) with different settings of integration time and light intensity and compared the relative changes as a function of wavelength -- see left panel of Fig.~\ref{fig:dusanerr}. This shows that the measurements agree within 10\% down to 365\,nm, where the differences get overwhelmed with statistical fluctuations as the transmissivity of the filter is negligible. The largest systematic effect on the calculation of the molecular extinction, due to its steep dependence on wavelength, comes from the shape of the short-wavelength slope of $R(\lambda)$. To estimate an upper bound on this uncertainty, we thus assume a non-realistic worst case scenario, where the measurement shifts by 10\% right at the position of this slope (see right panel of Fig.~\ref{fig:dusanerr}) -- even such a situation leads only to a shift of 0.001 in the the molecular extinction, showing that the precision of the methods used to determine $R(\lambda)$ is sufficient for this purpose. Similarly, assuming that the lens transmissivity does not decrease to zero at 365\,nm but follows exactly the (rather smooth) measured data at least down to 350\,nm changes the predicted molecular extinction only by 0.0007.

The largest uncertainty in the extinction model is that its parameters are determined from a set of stellar spectra which may not correspond to the actual stellar population in the area of measurement. From Fig.~\ref{fig:ray} we estimate the maximal resulting uncertainty in the determination of $\tau_\mathrm{FRAM}$ to be roughly 0.002 in the worst case scenario where the the distribution of the measured stars is heavily biased towards a specific part of the spectral library used. This uncertainty directly translates to an overall offset in the value of VAOD.

The uncorrelated systematic uncertainty should ultimately be summed in squares with the statistical uncertainty of individual fits, which is 0.001--0.002 for all scans in the test sample, leading to the final estimate of the maximal uncertainty of 0.006 correlated and 0.005 uncorrelated.

\begin{figure}
\begin{centering}
\includegraphics[width=\textwidth]{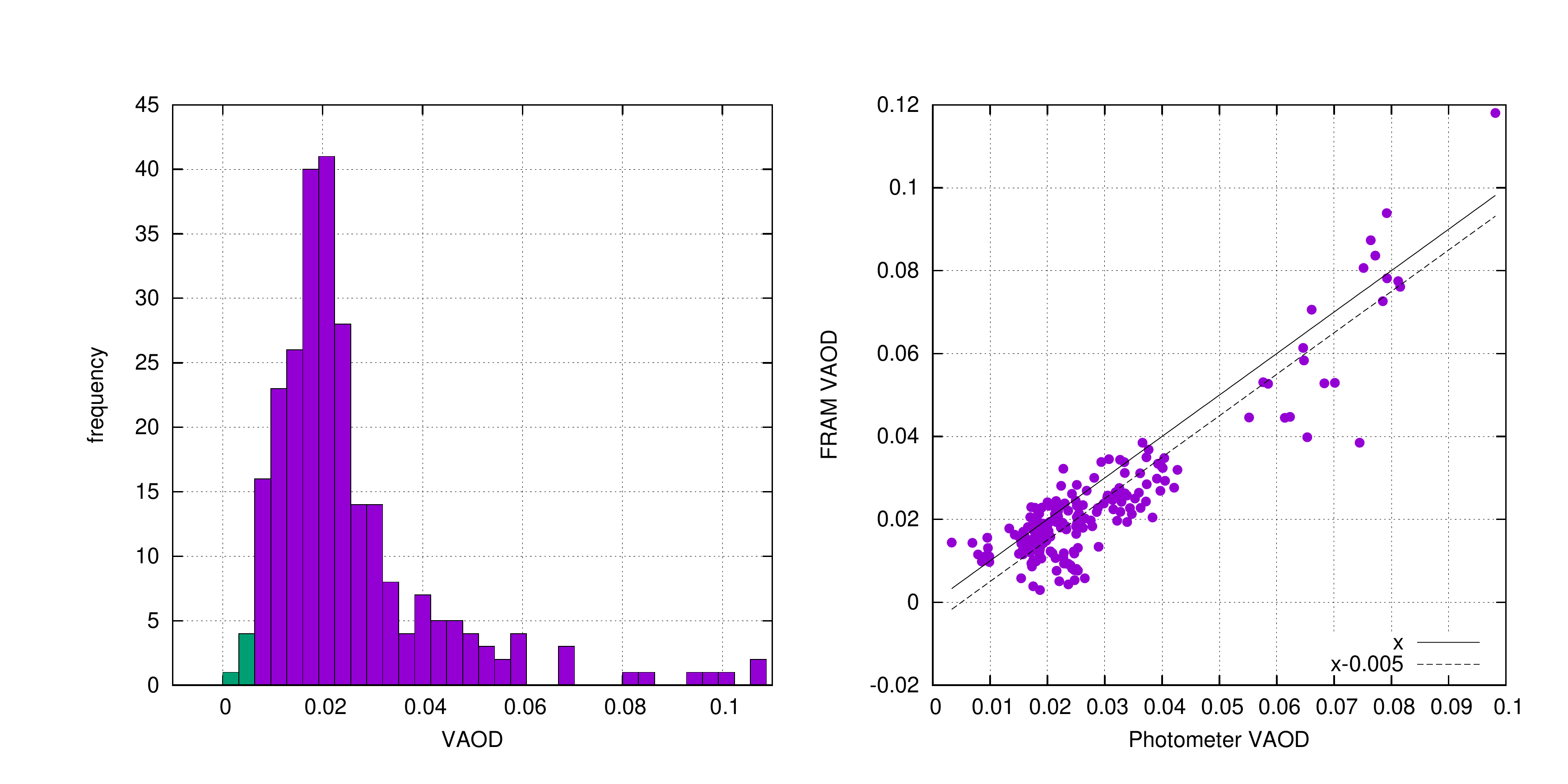}
\par\end{centering}
\caption{Left: histogram of VAOD values from the test sample of 263 scans from three different setups; values lower than the stratospheric contribution are highlighted in green (the binning has been chosen so that the value lies at a bin boundary). Right: comparison between the VAOD measured by the CTA-N FRAM and by a nearby Sun/Moon Photometer (in lunar mode), selected for observations less within 15 minutes apart; the $y=x$ line is added as reference as well as the dashed line shifted by the mean systematic difference between the values (0.005).}
\label{fig:lastfig}
\end{figure}

In Fig.~\ref{fig:lastfig} we present two plots that support such precision as realistic. The left panel shows the histogram of the VAOD values for the 263 test scans, starting sharply at the value of the stratospheric contribution (which is always present) -- this feature is particularly sensitive to the correct modeling of the molecular extinction, which for most values of VAOD measured constitutes the vast majority of atmospheric extinction at these wavelengths. In the right panel, we show the comparison of the FRAM VAOD values with those measured by a nearby Sun/Moon Photometer in lunar mode \citep{jurysek}. The mean systematic difference between the FRAM and the photometer is 0.005 with RMS of 0.007 -- consistent with the quoted FRAM uncertainties despite the photometer measurements having uncertainties of their own (albeit difficult to estimate), which indicates that, apart of possible effects that are correlated for both instruments, the FRAM uncertainty cannot be much larger.

\section{Conclusions and Outlook}
\label{concl}

We have shown that using wide-field photometry, we can measure the Vertical Atmospheric Optical Depth with a precision (for a single measurement) better than 0.008 optical depths. This is a relatively cheap and technically simple method, especially in environments where the production of laser light at night is undesired, such as astronomical observatories. Because the determination of the VAOD with this method is independent on the absolute calibration of the system, the method is suitable for long-term monitoring, as slow changes in detector response do not bias the results.

The method presented depends on the assumption of horizontal uniformity (stratification) of aerosols. On the selected ``clear'' scans, the fit gives values of $\chi^2$ per degree of freedom between 1.1--1.4, using the error model developed in Sec.~\ref{errm}, indicating very good description of the altitude dependence of extinction. However, this might be due to the fact that any scans performed during conditions of non-homogeneous aerosols is discarded as ``cloudy''. This is clearly an area demanding further study, which could be also facilitated through coordination with other atmospheric monitoring instruments that are or will soon be deployed near the future CTA sites.

\acknowledgments

This work was supported by European Structural and Investment Fund and the Czech Ministry of Education, Youth and Sports (MEYS) within the projects CZ.02.1.01/0.0/0.0/16\_013/0001403, CZ.02.1.01/0.0/0.0/18\_046/0016007 and CZ.02.1.01/0.0/0.0/15\_003/0000437) and by the MEYS projects LM2018105 and LTT17006. We appreciate access to computational resources provided by Computing center of the Institute of Physics of the Czech Academy of Sciences \citep{farma} and supported by project LG13007. This work was conducted in the context of the CTA Consortium. 

\software{IRAF \citep{iraf,iraf2}, SExtractor \citep{sextractor}, MODTRAN \citep{modtran}, Astrometry.net \citep{astrometry.net}}

%




\bibliography{ebr-aerosol-wf}{}
\bibliographystyle{aasjournal}



\end{document}